\documentclass[12pt]{article}
\usepackage{graphicx}
\begin{document}

\centerline{\Large\bf Soft spin dipole giant resonances in $^{40}$Ca}

\begin{center}
L. Stuhl$^1$, A. Krasznahorkay$^1$, M. Csatl\'os$^1$,
T. Marketin$^{2,3}$, E.~Litvinova$^{2,4}$, 
T.~Adachi$^5$, A. Algora$^6$,  J.~Daeven$^7$,  E. Estevez$^6$, 
H.~Fujita$^5$, 
Y. Fujita$^8$, C.~Guess$^7$,  J. Guly\'as$^1$,  
K. Hatanaka$^5$, 
K.~Hirota$^5$, H.J. Ong$^5$,  D. Ishikawa$^5$, H.~Matsubara$^5$, 
R.~Meharchand$^7$, F.~Molina$^6$,   H. Okamura$^5$, 
G.~Perdikakis$^7$,  B.~Rubio$^6$, C. Scholl$^9$,  T.~Suzuki$^5$,
G. Susoy$^{10}$,  A.~Tamii$^5$,  J. Thies$^{11}$, R.~Zegers$^7$, J.~Zenihiro$^5$ \\
 \vskip 0.5cm
$^1$  Inst. of Nucl. Res. (ATOMKI), P.O. Box 51, H-4001 Debrecen, Hungary, \\
$^2$ GSI 64291 Darmstadt, Germany \\
$^3$ Phys. Dept. Univ. of Zabreb, 10000 Zagreb, Croatia \\
$^4$ Inst. f\"ur Theor. Phys. Goethe Univ., 60438 Frankfurt am Main, Germany \\
$^5$  RCNP, Osaka University, Ibaraki, Osaka 567-0047, Japan\\
$^6$ IFIC, CSIC-Universidad de Valencia, 46071 Valencia, Spain\\
$^7$ NSCL, Michigen State University, East Lansing, Michigan 48824, USA\\
$^8$ Department of Physics, Osaka University, Toyonaka, Osaka 560-0043, Japan\\
$^9$ Institut für Kernphysik, Universität zu Köln, 50937 Köln, Germany\\
$^{10}$ Istanbul Univ., Faculty of Sci., Phys. Dept., 34134 Vezneciler, Istanbul, 
Turkey\\
$^{11}$ Institut für Kernphysik, Universität Münster, 48149 Münster, Germany
\end{center}

\begin{abstract}

High   resolution    experimental   data    has   been   obtained    for   the \\
$^{40,42,44,48}$Ca($^3$He,t)Sc charge exchange reaction  at 420 MeV beam
energy, which
favors the  spin-isospin excitations. The measured  angular distributions were
analyzed for each state separately,  and the relative spin dipole strength has
been  extracted  for  the  first  time.  The  low-lying  spin-dipole  strength
distribution  in  $^{40}$Sc  shows  some  interesting periodic  gross
feature.  It resembles to a soft, dumped multi-phonon vibrational band with
$\hbar\omega$= 1.8 MeV, which might be associated to pairing vibrations around
$^{40}$Ca.
 

\end{abstract}

\section{Introduction}

It was realized already by A. Bohr \cite{bo69}, who discussed the
general properties  expected for a  full (nn, pp  and np) components  that T=1
pairing phonon  is appropriate for the  region around $^{40}$Ca.  
Already then, the
evidence  including both  energetics  and transfer  data. \cite{na69}  was  
compelling
regarding the major role played by  the T=1 pairing in N=Z nuclei. These ideas
were  further developed  by B\'es  and Broglia  and culminated  in  the review
article \cite{be77},  where a very  detailed analysis of isovector  pairing 
vibrations
was  presented.

Studies involving isospin effects have  undergone a resurgence in recent years
as such nuclei  become more readily accessible. Moreover,  near closed shells,
the   strength  of  the   pairing  force   relative  to   the  single-particle
level-spacing is expected to be less  than the critical value needed to obtain
a  superconducting  solution, and  the  pairing field  then  gives  rise to  a
collective phonon.  However, despite  many experimental efforts,
these predictions have not been confirmed yet.

Macchiavelly et al., \cite{mac2000} presented an experimental
analysis of the pairing vibrations around $^{56}$Ni with emphasis on odd-odd
nuclei. Their results clearly indicate a collective behavior of the isovector
pairing vibrations. The $\hbar\omega$  of such  vibrations is  estimated  to
be  0.8 MeV \cite{mac2000}  for $^{40}$Ca.
In  a recently  published  article  Cedervall   et  al., \cite{ceder2011}  
obtained evidence  for  a spin-aligned neutron-proton paired phase from the 
level structure of $^{92}$Pd. Gezerlis et al., \cite{gez2011} discussed also 
the possibility of mixed-spin
pairing correlations in heavy nuclei.
  
The aim of the present work was to study the low-lying dipole strength
distribution in $^{40}$Ca by using the ($^3$He,t) reaction, which was
extensively used earlier for the excitation of spin-isospin vibrational states
in other isotopes. Some of our preliminary results was published
recently \cite{stuhl2011}.

The $^{40}$Ca($^3$He,t)$^{40}$Sc reaction has been studied earlier by
 Schulz et al., \cite{schulz1970} at 28 MeV bombarding energy and by 
Loiseaux et al., \cite{loise71} at 30.2 MeV
 bombarding energy  using solid state
 telescopes and a magnetic spectrograph. The energy resolution was 70 keV and 
15-20
 keV, respectively. Some of the $(\pi 1f_{7/2})(\nu 1d_{3/2})^{-1}$ , 
$(\pi 1p_{3/2})(\nu 1d_{3/2})^{-1}$ 
and $(\pi 1f_{7/2})(\nu 2s_{1/2})^{-1}$  proton-neutron multiplet states are 
identified and the effect
of configuration mixing is discussed. A similar experiment has been performed
more recently  by Hansper et al., \cite{hans2000} at 26.1 MeV, using a magnetic
  spectrometer with an energy resolution of 15 keV. Correspondence of the
  observed  $^{40}$Sc levels with the known T=1 states in $^{40}$K and
  $^{40}$Ca are based on predictions provided by the isobaric multiplet mass
  equation.

The  $^{48}$Ca($^3$He,t)$^{48}$Sc  reaction was studied earlier by Grewe et al.,\cite{grewe2007} at 420 MeV
bombarding energy with an energy resolution of about 40 keV.
Up to about 9 MeV some  excited states relevant for the double $\beta$-decay
were identified. Those levels were observed
and investigated also in our present work and they are in good agreement with
the levels, observed by Grewe et al.

 The spin-isospin excitation has been
 investigated earlier by Tabor et al., \cite{tabor1982} in the
 $^{40}$Ca($^3$He,t)$^{40}$Sc 
reaction at 130 and 170 MeV. The angular distribution was
 measured for the suspected giant dipole resonance (GDR) structure. The data are
 reasonably well described by a collective model calculation based on the
 Goldhaber-Teller model of the GDR. Some weaker L=1 resonances at 2, 4, 6 and 8 MeV
 has also been observed. However, their energy resolution of about 400 keV did
 not allow to study their structures.

\section{Experimental methods and results}

The experiment was performed at the Research Center for Nuclear Phy\-sics, Osaka
University.  The energy of the $^3$He beam of $420$ MeV was achromatically
transported to self supporting metallic $^{40}$Ca ,$^{42}$Ca, $^{44}$Ca,
$^{48}$Ca targets with thicknesses of $1.63 - 1.87$ mg/cm$^2$. The typical beam
current was $5$ nA. The energy of the tritons was measured with a magnetic
spectrometer "Grand Riden", using complete dispersion matching techniques 
\cite{fujita}.
The energy resolution was about $20$ keV.  The spectrometer was set at 0$^\circ$
and 2.5$^\circ$ with respect to the beam direction with an opening angle of 
$\pm$20 mrad horizontally and $\pm$20
mrad vertically defined by a slit at the entrance of the spectrometer.
A few typical triton spectrum is shown in the left side of Fig. 1.

\begin{figure}
\begin{center} 
\includegraphics[width=140mm]{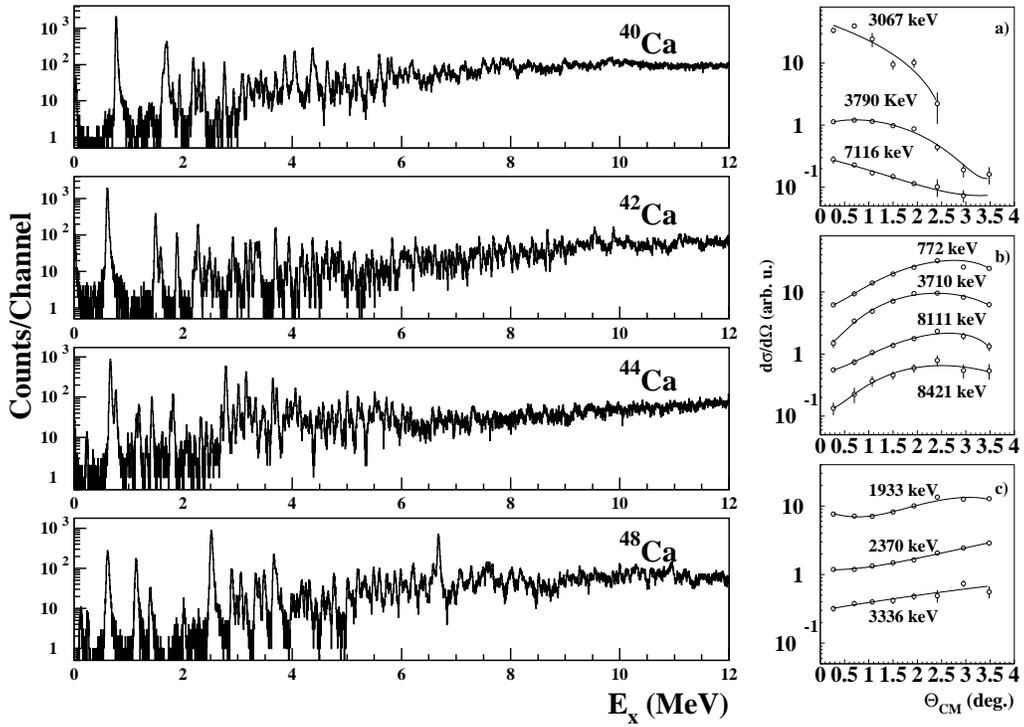}
\end{center}
\caption {Left side: Part of  the triton  spectra measured  
at $\theta= 2.5^\circ$ for the different Ca targets indicated in the 
figures, showing preferentially the $\Delta$L=1, 
dipole transitions. Right side: examples for the angular distributions
measured 
in the $^{40}$Ca($^3$He,t) $^{40}$Sc reaction 
for  $\Delta$L=0 (a), for $\Delta$L=1 (b) and for some not identified (c) 
transitions}.
\end{figure}

The spectra were analyzed using the program package: Gaspan. In a given energy 
range all 
peaks were fitted at the same time. 
Gaussian line shape with exponential tails and  second order polynomials were 
used for
describing the background. The quality of the fit was always good. The excitation
energies of the IAS's and a few well known excited states are used for
determining the precise energy calibration. We determined the precise level
energies and intensities for each isotope. The spectra were
studied in eight distinct angular regions for all scandium isotopes. The
angular distributions were determined for each known, and new peaks. A few
examples for the angular distributions are displayed at the right hand side of
Fig. 1.

The lowest lying states in $^{40}$Sc are identified as members of the
 $(\pi 1f_{7/2})$ $(\nu 1d_{3/2})^{-1}$ (J$^\pi$(J$^\pi$ = $2^- - 5^-$) multiplet. 
As the ($^3$He,t) reaction at this bombarding energy excites preferentially
the spin-flip states, in our case the 2$^-$ state is excited the strongest,  
in which the spin of the proton and the neutron hole is parallel. 
The isospin of such state is T=1 and T$_z$=-1.

Such proton neutron multiplet has been observed also in the mirror nucleus of
$^{40}$Sc namely in $^{40}$K in which it is also the ground state multiplet with
 T=1 and T$_z$=1.
The T$_z$=0 members of the isospin multiplet has been observed in $^{40}$Ca at
7.658 MeV above the ground state. Such a shift can be explained by the Coulomb 
energy difference.
The other states of the proton-neutron multiplet with J$^\pi$= 3$^-$, 4$^-$
and 5$^-$ were also identified in  $^{40}$Sc(T=-1), in $^{40}$Ca(T=0) and 
in $^{40}$K(T=1) \cite{nds}. 

All three multiplets turned out to be very similar. The sequence of the
J$^\pi$'s 
are 4$^-$, 3$^-$, 2$^-$ and
5$^-$. The energy differences between the members of the bands agree(s) within 
1-2 keV.

In $^{40}$Sc no other multiplet states has been identified yet. 
However, using the strong similarity of the low-lying excited states in  
$^{40}$K and  $^{40}$Sc we may identify some additional multiplet states.
The next multiplet in $^{40}$K is  $(\pi 1d_{3/2}))^{-1}(\nu 1d_{3/2})^{-1}$ 
with L=2 and J$^\pi$'s are 0$^+$ (1644 keV), 2$^+$ (1959 keV),  3$^+$ (2260 keV)
 and 1$^+$ (2290 keV).

At about the same excitation energy the $(\pi 1d_{3/2}))^{-1}(\nu 2p_{3/2})$ 
(L=1, J$^\pi$ = $0^- - 3^-$) multiplet was also identified. As the  $(\pi 2p_{3/2})$
is somewhat lower in  $^{41}$Sc than in $^{39}$K such a multiplet should be also
lower in $^{40}$Sc than in $^{40}$K. The triplet state observed around 
1.7 MeV is a good candidate for the $(\nu 1d_{3/2}))^{-1}(\pi 2p_{3/2})$ 
multiplet in $^{40}$Sc. This assignment is supported also by the angular 
distribution of the states.

The dipole strengths distributions deduced from our experimental data 
for $^{40}$Sc, 
$^{42}$Sc, $^{44}$Sc, and $^{48}$Sc are compared in Fig. 2.

\begin{figure}
\centering \includegraphics[width=120mm]{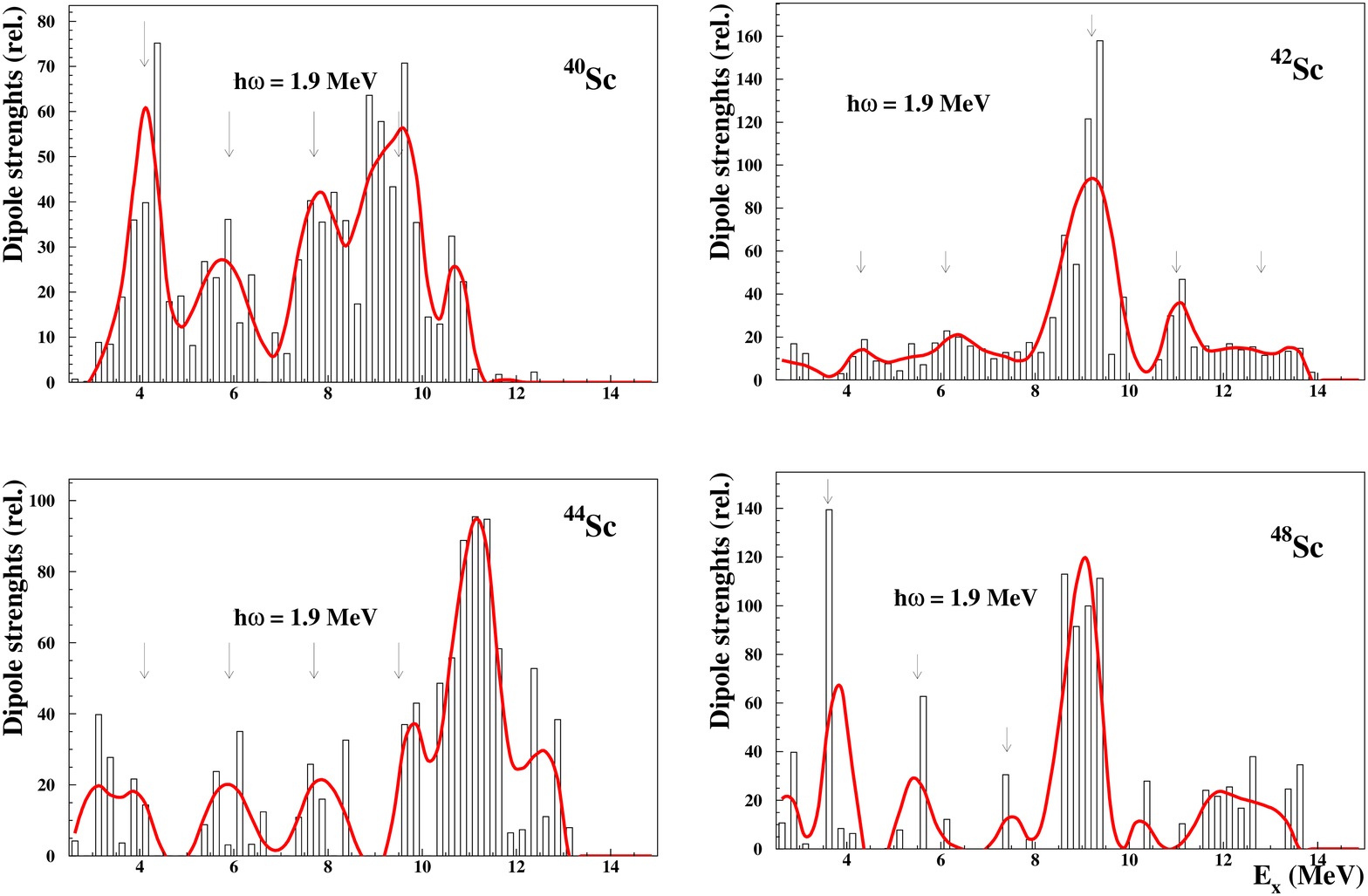}
\caption  [] {Relative dipole strengths distributions for $^{40}$Sc (a), 
$^{42}$Sc (b), $^{44}$Sc (c), and $^{48}$Sc (d) as a function of the
  excitation energy.}
\end{figure}

\section{Conclusions}

Concerning the dipole strengths distributions shown in Fig. 2 two observation
 can be made: 
\begin{enumerate}
\item{} there is a 
relatively strong peak at about 10 MeV in each distributions,

\item{}  some periodic
structure of the distribution is showing up, especially for  $^{40}$Sc.

Both features are very common for all isotopes, which suggests the presence of
some core excitations in $^{40}$Ca.
\end{enumerate}

Both features are very common for all isotopes, which suggests the presence of
some core excitations in $^{40}$Ca.

In order to understand the experimental results relativistic RPA (RRPA)
calculations have been performed with NL3 \cite{lala97} and DD-ME2 \cite{lala05} interactions. The
results are shown in Fig. 3. RRPA predicts low-lying strength with nearly
periodic peaked structure caused by the dipole isospin-flip and
spin-isospin-flip transitions governed by the pion and effective rho-meson
exchange interactions. The obtained strength, however, shows no enhancement of
the strength at lowest energies. Thus, it is expected that correlations beyond 
RRPA can be responsible for the observed enhancement.

\begin{figure}[ht]
\begin{center} 
\includegraphics[width=120mm]{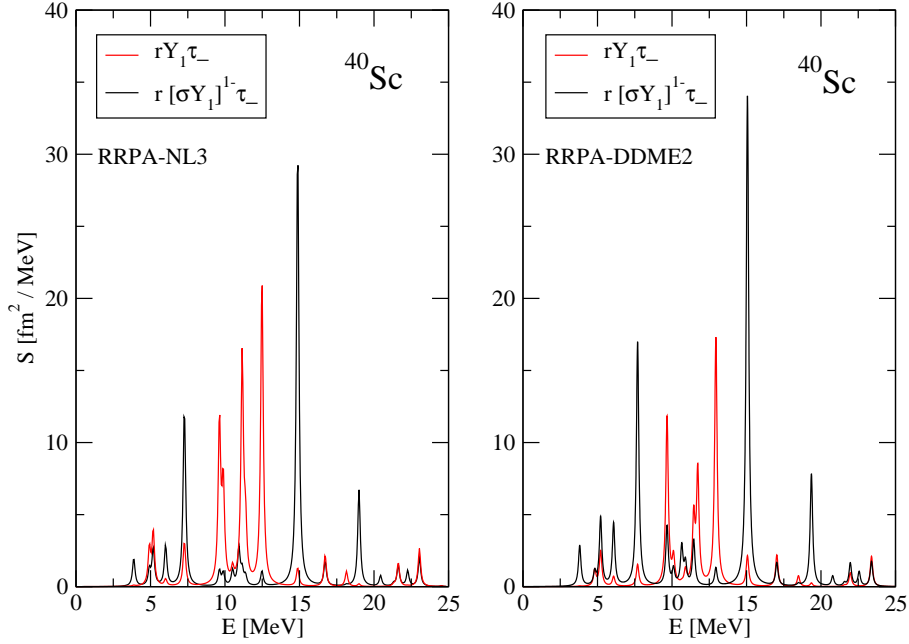}
\caption  [] {Isospin-flip and spin-isospin-flip dipole strength functions obtained within the
 Relativistic RPA with NL3 and DD-ME2 interactions.}
\end{center}
\end{figure}

It was realized that the periodic structure observed in the dipole strengths
distribution can be associated with the multiparticle-multihole $0^+$ states
observed previously in $^{40}$Ca, and also shown in the left part of Fig. 4.

\begin{figure}[htb]
\begin{center} 
\includegraphics[width=120mm]{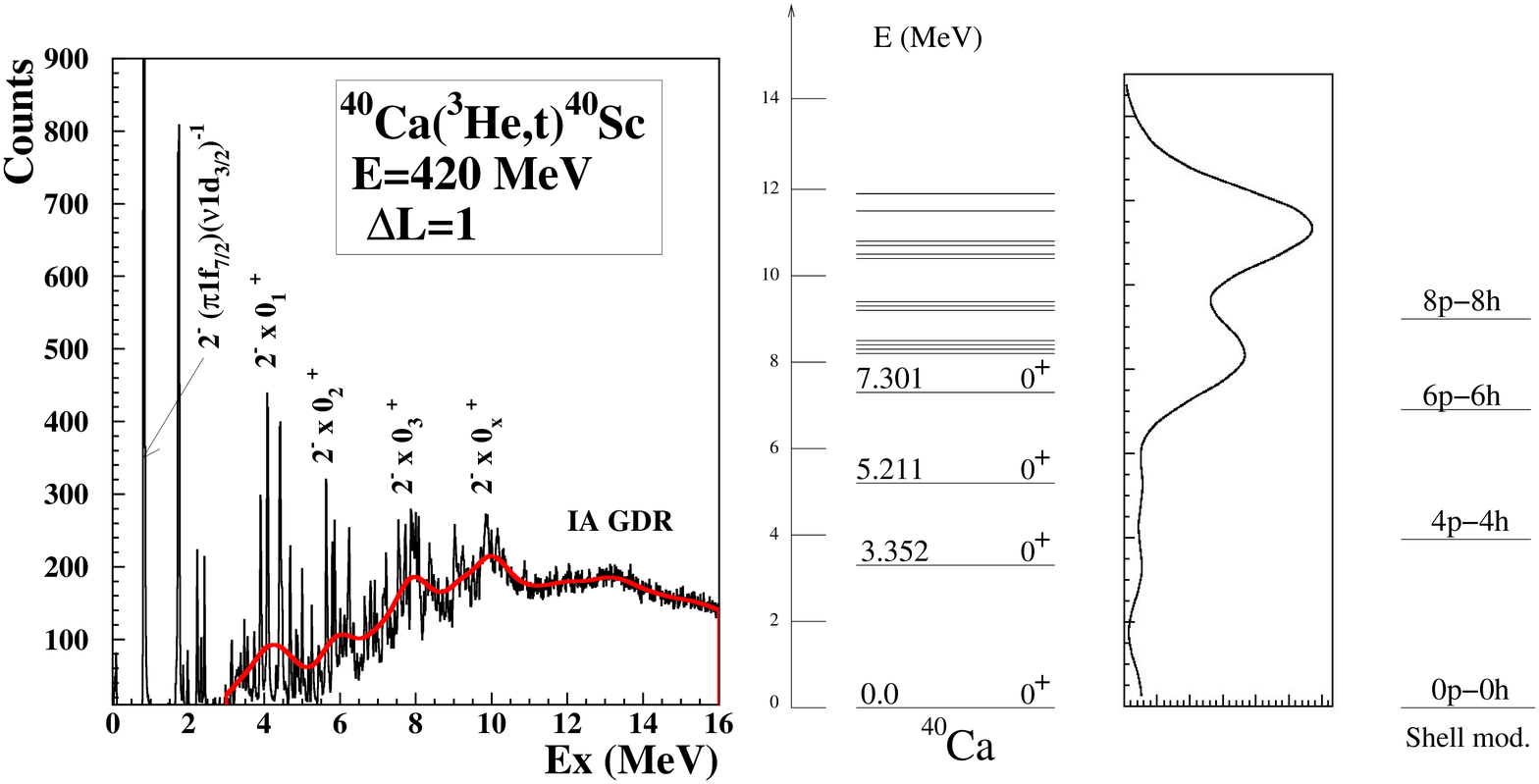}
\caption  [] {Left part: Spin-dipole excited states observed in the $^{40}$Ca($^3$He,t)Sc
  reaction at $\Theta =
  2.5^\circ$. A folded spectrum (with FWHM=500 keV) is shown in red. The
  periodic structure of the distribution is clearly visible.  
The position of the strong 2$^-$ transition of the ground state multiplet as
well as the results of their coupling to the lowest lying 0$^+$-states in
$^{40}$Ca are marked.
Right part: Part of the level scheme of $^{40}$Ca showing only the 0$^+$
  levels, the level density of the 0$^+$-states and the shell model prediction
  for the multiparticle-multihole configurations \cite{saka}.}
\end{center}
\end{figure}

Coupling the 2$^-$ state of the ground state multiplet (which is the strongest
channel in the ($^3$He,t) rection) to the different low-lying 0$^+$ states in
$^{40}$Ca, the centroids of the bumps at 4,6 and 8 MeV can nicely be
reproduced. 
The density of the states increases rapidly above 8 MeV. The
distribution of the 0$^+$ states is also shown in the right part of the
figure. It has a definite peak at around 10 MeV. Coupling this peak to
the 2$^-$ state, one gets the 10 MeV peak in the SDR distribution.

Such low-lying  0$^+$ states in $^{40}$Ca can also be  considered as parts of 
the Giant
Monopole Resonance (GMR). The GMR was investigated by Yangblood et al.,
\cite{yang1997} and indeed observed a bump around 10 MeV, although the
centroid of the GMR was found to be at 19 MeV.

We expect also coupling
of the 2$^-$ state  to the GMR. The high energy dipole stench distribution was
investigated by Gaarde et al. \cite{ga1981} in the  $^{40}$Ca(p,n) reaction at
200 MeV bombarding energy and observed such a strong dipole peak at about 22 MeV. 

According to
the shell model calculations of M. Sakakura et al.,\cite{saka} the energy of the 4,6 and 8
particle-hole states is also shown in Fig. 4.

Such multiparticle-multihole configurations might be associated to monopole
multiphonon states as well. We have a proton-neutron pair connected to such 
multiphonon
states. The resulting 2$^-$ states are also coupled to hundreds of other 2$^-$
states (we are dealing with an odd-odd nucleus), which result in the observed
spreading of their strengths.

Similar periodic structure is expected in the dipole strengths distribution
of $^{40}$K excited in the $^{40}$Ca(n,p) reaction.
For the N=Z $^{40}$Ca nucleus the sum rule for the $\beta^-$ and $\beta^+$
strengths reduces to $S_{SDR}^- = S_{SDR}^+$ \cite{kraszna99}.
So the cross section of the  $^{40}$Ca(p,n) and  the $^{40}$Ca(n,p)  reactions 
should be
the same. For isospin symmetry considerations one expects the same strengths
distribution  as well for the (p,n) and (n,p) type reactions.

Unfortunately, most of the (n,p) reactions were performed at low bombarding 
energy
where the spin-isospin exciations were suppressed. The only reaction, which
was studied at intermediate energy (152 MeV) was performed by Maesday et al.,
\cite{maes}, but the energy resolution was about 3 MeV, which smeared out the
structure of the spin dipole strengths distribution. They observed only one
broad peak, which was identified as the GDR.

\

\noindent{\bf Acknowledgments}

The authors acknowledge the RCNP cyclotron staff for their support during the
course of the present experiment.
This work has been supported  by the
Hungarian OTKA Foundation No.\, K106035 and by the LOEWE program of the State 
of Hesse (Helmholtz International Center for FAIR).

\end{document}